\documentclass[letter]{aa}

\usepackage{txfonts}
\usepackage{graphicx}
\usepackage{natbib}

\begin{document}

\title{X-ray emission from MP~Muscae: an old classical T Tauri star}

\author{C.~Argiroffi\inst{1} \and A.~Maggio\inst{2} \and G.~Peres\inst{1}}
\offprints{C.~Argiroffi, {\email argi@astropa.unipa.it}}
\institute{Dipartimento di Scienze Fisiche ed Astronomiche, Sezione di Astronomia, Universit\`a di Palermo, Piazza del Parlamento 1, 90134 Palermo, Italy, \email{argi@astropa.unipa.it, peres@astropa.unipa.it} \and INAF - Osservatorio Astronomico di Palermo, Piazza del Parlamento 1, 90134 Palermo, Italy, \email{maggio@astropa.unipa.it}}
\date{Received 22 December 2006 / Accepted 25 January 2007}

\titlerunning{X-ray emission from the old CTTS MP~Mus}
\authorrunning{C.~Argiroffi et al.}

\abstract
{}
{We study the properties of X-ray emitting plasma of MP~Mus, an old classical T~Tauri star. We aim at checking whether an accretion process produces the observed X-ray emission and at deriving the accretion parameters and the characteristics of the shock-heated plasma. We compare the properties of MP~Mus with those of younger classical T~Tauri stars to test whether age is related to the properties of the X-ray emission plasma.}
{XMM-Newton X-ray spectra allows us to measure plasma temperatures, abundances, and electron density. In particular the density of cool plasma probes whether X-ray emission is produced by plasma heated in the accretion process.}
{X-ray emission from MP~Mus originates from high density cool plasma but a hot flaring component is also present, suggesting that both coronal magnetic activity and accretion contribute to the observed X-ray emission. We find a Ne/O ratio similar to that observed in the much younger classical T~Tauri star BP~Tau. From the soft part of the X-ray emission, mostly produced by plasma heated in the accretion shock, we derive a mass accretion rate of $5\times10^{-11}\,{\rm M_{\sun}\,yr^{-1}}$.}
{}


\maketitle

\section{Introduction}

Low mass stars are sources of strong X-ray radiation since their early evolutionary phases. Coronal plasma, responsible for the X-ray emission, is confined and probably heated by magnetic fields which emerge from the stellar surface \citep{FeigelsonMontmerle1999,PreibischKim2005}. The coronal plasma observed in essentially all late-type stars can be characterized by a large variety of average temperatures (from few MK to tens of MK) and metallicities (from one tenth to few times the solar photospheric value), but a common feature is the low density measured at the temperature of formation of \ion{O}{vii} He-like triplets ($N_{\rm e} \approx 10^{10}-10^{11}$\,cm$^{-3}$ at $T \sim 2$\,MK) \citep{TestaDrake2004,NessGudel2004}.

In very young stars, however, accretion may cause X-ray emission in addition to magnetically-confined coronal plasma. In classical T~Tauri stars (CTTSs) gas falls from the circumstellar envelope, funnelled by the magnetic field, and hits the stellar photosphere. In the resulting shock the accreted material is heated to temperatures of few MK \citep{CalvetGullbring1998}. With typical mass accretion rates, the shock-heated plasma can reach X-ray luminosities as high as $10^{31}\,{\rm erg\,s^{-1}}$.

The few CTTSs for which high-resolution X-ray spectroscopy was performed up to date show, in most cases, cool plasma components ($2-4$\,MK) with large electron densities ($10^{11}-10^{13}$\,cm$^{-3}$) which have been interpreted as evidence for X-ray emission due to an accretion shock. The best known examples of this behavior are the CTTSs TW~Hya, BP~Tau, and V4046~Sgr \citep{KastnerHuenemoerder2002,SchmittRobrade2005,GuntherLiefke2006}. Noticeable exceptions are the CTTS T~Tau \citep{GudelSkinner2007} and the Herbig star AB~Aur \citep{TelleschiGudel2007a}.

In the hypothesis of X-ray emission originated in shocks great attention has been focused also on plasma element abundances. In fact, they probe the chemical composition of the accreting stream, and hence provide insightful indications on the physical and chemical processes at work in the inner circumstellar disk \citep{StelzerSchmitt2004,DrakeTesta2005b}.

In this letter we present the {\it XMM-Newton} observation of MP~Muscae, one of the oldest known CTTSs, aimed at studying the properties of the X-ray emitting plasma and the role of the accretion process. MP~Mus is a K1~IVe star of the Lower Centaurus Crux (LCC) association. \citet{GregorioHetemLepine1992} first identified it as a classical T~Tauri star by measuring enhanced H$\alpha$ emission (${\rm EW}=-47$\,\AA) and Li absorption (${\rm EW}=0.37$\,\AA). Excesses in infrared bands revealed an optically thick circumstellar disk \citep{MamajekMeyer2002,SilverstoneMeyer2006} with an estimated dust mass of $\sim5\times10^{-5}\,{\rm M_{\sun}}$ \citep{CarpenterWolf2005}. \citet{BatalhaQuast1998} derived a rotational period of 5.75\,d from variability in the B, V, R, and I bands, although the amplitude of variability is surprisingly low for a CTTS.

\citet{MamajekMeyer2002} derived three different ages for MP~Mus, 7, 14, and 17\,Myr, depending on the adopted theoretical evolutionary tracks. The LCC association, to which MP~Mus belongs, is the oldest portion of the Scorpius-Centaurus OB~association, and its estimated age is between 16 and 23\,Myr \citep{MamajekMeyer2002,SartoriLepine2003}.

\begin{figure} 
\resizebox{\hsize}{!}{\includegraphics{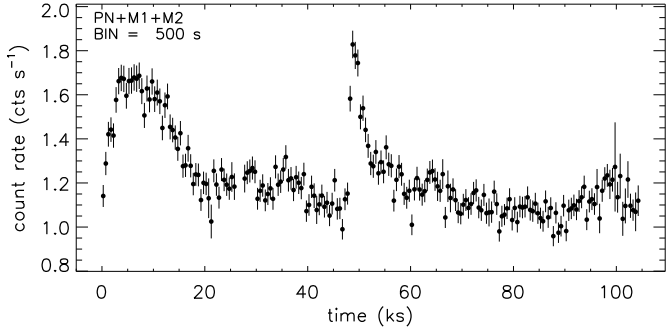}}
\caption{Background-subtracted light curve of MP~Mus obtained by adding the three EPIC instruments.}
\label{fig:lc}
\end{figure}

\section{Observation and data analysis} \label{obs}

MP~Mus was observed with {\it XMM-Newton} for a duration of $\sim110$\,ks on 2006 August 19--20. We processed the data using the SAS V6.5 standard tasks. After having discarded time segments affected by high background count rates, we obtained good time intervals summing up to $\sim100$\,ks. The X-ray light curve (Fig.~\ref{fig:lc}) shows clear hints of flaring activity, typical of magnetically active coronal sources.

To increase the signal-to-noise ratio we rebinned the PN and MOS spectra to obtain at least 30 counts in each bin, and the RGS wavelength bins were joined two by two. Spectral analysis was performed using the Astrophysical Plasma Emission Database \citep[APED V1.3,][]{SmithBrickhouse2001}. We adopted the \citet{AsplundGrevesse2005} solar photospheric composition as the reference set of element abundances.

We derived the characteristics of the X-ray emitting plasma of MP~Mus by fitting PN and MOS spectra with XSPEC V11.3.0 \citep{Arnaud1996}, adopting an optically-thin plasma emission model with three isothermal components. The best-fit 3-T model (Table~\ref{tab:res}) also provided individual O, Ne, Fe, and S abundances, while the abundances of the other elements were tied to the Fe abundance because the fit quality does not improve if they are treated as additional free parameters.

\begin{table}
\caption{MP~Mus best-fit paramenters.}
\label{tab:res}
\begin{center}
\begin{tabular}{lcccc}
\hline
\hline
\multicolumn{1}{l}{Par.}     & \multicolumn{4}{c}{best-fit value} \\
\hline
T$^a$         & $2.7^{+0.1}_{-0.2}$ & $7.2^{+0.4}_{-0.5}$ & $36^{+18}_{-11}$     & \\
EM$^b$        & $9.6^{+5.2}_{-2.5}$ & $17.9^{+4.0}_{-3.0}$ & $2.9^{+1.5}_{-0.8}$ & \\
$N_{\rm H}^c$ & $4.6^{+1.8}_{-1.7}$ & & & \\
Ab.$^d$       & O=$0.25^{+0.08}_{-0.07}$ & Ne=$0.76^{+0.23}_{-0.15}$ & S=$0.28\pm0.20$ & Fe=$0.09^{+0.04}_{-0.02}$ \\
\hline
\end{tabular}
\end{center}
$^a$~Temperature (MK). $^b$~Emission Measure $({\rm10^{52}\,cm^{-3}})$. $^c$~Hydrogen column density $({\rm 10^{20}\,cm^{-2}})$. $^d$~Abundances referred to the solar photospheric values of \citet{AsplundGrevesse2005}. All the uncertainties correspond to the 68\% confidence level.
\end{table}

Table~\ref{tab:lines} shows the fluxes of the strongest RGS lines, measured with PINTofALE V2.0 \citep{KashyapDrake2000}, adopting a lorentzian function to fit the line profile. We checked that the derived 3-$T$ EPIC model provides a reasonably good description of the RGS line fluxes.

\begin{figure} 
\resizebox{\hsize}{!}{\includegraphics{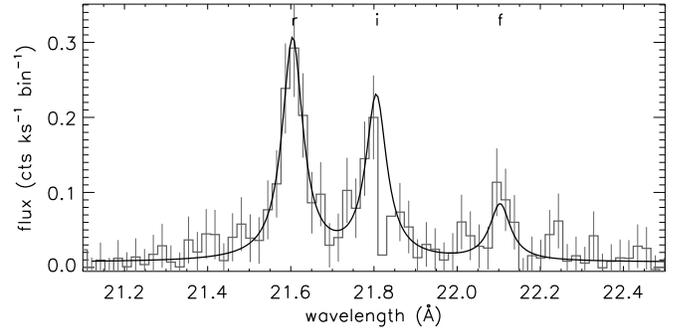}}
\caption{RGS1 spectrum in the wavelength region of the \ion{O}{vii} triplet (gray) with the best fit lorentzian line profile (black). Wavelength bins corresponding to bad column pixels are plotted without error bars and were not considered in the fitting procedure.}
\label{fig:ovii}
\end{figure}

\begin{figure}
\resizebox{\hsize}{!}{\includegraphics{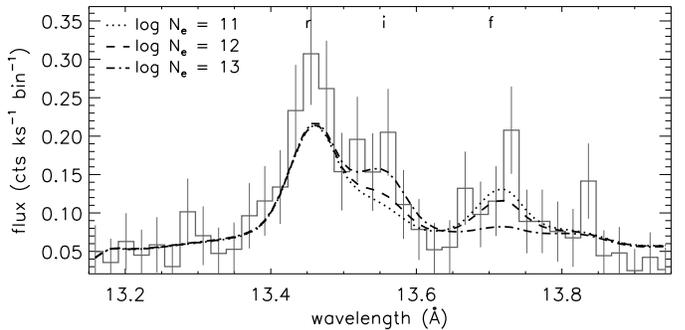}}
\caption{RGS2 spectrum around the \ion{Ne}{ix} triplet (gray) with the predicted spectra obtained from the 3-$T$ EPIC model and assuming different electron densities.}
\label{fig:neix}
\end{figure}

We identified the \ion{O}{vii} and \ion{Ne}{ix} He-like triplets in the RGS spectra. The \ion{O}{vii} lines (indicated as $r$, $i$, and $f$ in Fig.~\ref{fig:ovii}) provide a density-sensitive ratio $f/i = 0.28\pm0.13$, which implies an electron density $\log N_{\rm e}=11.7\pm0.2$ for the plasma at $T\sim2$\,MK. The \ion{Ne}{ix} line flux measurements are affected by large uncertainties due to the strong blending with Fe lines. Figure~\ref{fig:neix} shows the observed \ion{Ne}{ix} triplet with that predicted on the basis of the 3-$T$ EPIC model assuming different $N_{\rm e}$ values. This comparison suggests that the plasma at $T\sim4$\,MK, which produces the \ion{Ne}{ix} triplet, has a density $\log N_{\rm e} > 11$, and possibly as high as that estimated from the \ion{O}{vii}.

\section{Discussion}

From the analysis of the \ion{O}{vii} triplet we find that the cool plasma component in our source has a density significantly larger than typical coronal values. It suggests that shock-heated plasma contributes significantly to the observed X-ray emission. In this respect, MP~Mus is the fourth CTTS discovered so far with evidence of X-ray emission produced by cool high density plasma, likely resulting from an accretion process. Previous cases were TW~Hya, BP~Tau, and V4046~Sgr. Moreover, MP~Mus shows also clear evidence of intense coronal activity, as indicated by the flares (see Fig.~\ref{fig:lc}) and by the hot plasma component.

\begin{table}
\caption{Strongest RGS lines of MP~Mus.}
\label{tab:lines}
\tiny
\begin{center}
\begin{tabular}{rrlcr@{$\pm$}l}
\hline\hline
 \multicolumn{1}{c}{$\lambda_{\rm obs}^{a}$} & \multicolumn{1}{c}{$\lambda_{\rm pred}^{a}$} & \multicolumn{1}{c}{Ion} & $\log T_{\rm max}^{b}$ & \multicolumn{2}{c}{$(F \pm \sigma F)^{c}$} \\
\hline
            12.14 &             12.13 &                      \ion{Ne}{x} \ion{Ne}{x} \ion{Fe}{xvii}  &   6.80 &                                         22.2 &   3.4 \\
            12.30 &             12.28 &                                \ion{Fe}{xxi} \ion{Fe}{xvii}  &   7.00 &                                          5.5 &   2.6 \\
            12.86 &             12.85 &         \ion{Fe}{xx} \ion{Fe}{xx} \ion{Fe}{xx} \ion{Fe}{xx}  &   7.00 &                                          3.4 &   2.3 \\
            13.46 &             13.45 &                                  \ion{Ne}{ix} \ion{Fe}{xix}  &   6.60 &                                         21.1 &   3.7 \\
            13.55 &             13.52 &                                  \ion{Ne}{ix} \ion{Fe}{xix}  &   6.90 &                                          8.6 &   3.1 \\
            13.73 &             13.70 &                                                \ion{Ne}{ix}  &   6.60 &                                          8.8 &   2.8 \\
            14.19 &             14.21 &                             \ion{Fe}{xviii} \ion{Fe}{xviii}  &   6.90 &                                          3.3 &   1.7 \\
            15.02 &             15.01 &                                              \ion{Fe}{xvii}  &   6.70 &                                         11.8 &   2.9 \\
            15.21 &             15.18 &    \ion{O}{viii} \ion{O}{viii} \ion{Fe}{xix} \ion{Fe}{xvii}  &   6.50 &                                          7.6 &   2.5 \\
            16.02 &             16.01 & \ion{O}{viii} \ion{O}{viii} \ion{Fe}{xviii} \ion{Fe}{xviii}  &   6.50 &                                         22.6 &   6.3 \\
            16.78 &             16.78 &                                              \ion{Fe}{xvii}  &   6.70 &                                          8.8 &   1.8 \\
            17.08 &             17.05 &                               \ion{Fe}{xvii} \ion{Fe}{xvii}  &   6.70 &                                         17.2 &   3.0 \\
            18.65 &             18.63 &                                                \ion{O}{vii}  &   6.30 &                                          8.7 &   2.6 \\
            18.98 &             18.97 &                                 \ion{O}{viii} \ion{O}{viii}  &   6.50 &                                         65.6 &   4.7 \\
            21.60 &             21.60 &                                                \ion{O}{vii}  &   6.30 &                                         30.2 &   4.0 \\
            21.81 &             21.80 &                                                \ion{O}{vii}  &   6.30 &                                         27.9 &   7.9 \\
            22.10 &             22.10 &                                                \ion{O}{vii}  &   6.30 &                                          8.0 &   2.8 \\
            24.80 &             24.78 &                                   \ion{N}{vii} \ion{N}{vii}  &   6.30 &                                         10.5 &   2.8 \\
            28.49 &             28.47 &                                     \ion{C}{vi} \ion{C}{vi}  &   6.20 &                                          3.8 &   1.3 \\
            33.74 &             33.73 &                                     \ion{C}{vi} \ion{C}{vi}  &   6.10 &                                         16.4 &   5.2 \\
\hline
\end{tabular}
\end{center}
$^a$~Observed and predicted (APED database) wavelengths (\AA).
$^b$~Temperature (K) of maximum emissivity.
$^c$~Line fluxes (${\rm 10^{-6}\,ph\,s^{-1}\,cm^{-2}}$) with uncertainties at the 68\% confidence level.
\normalsize
\end{table}
\normalsize

\subsection{Abundances}

The X-ray emitting plasma in MP~Mus is heavily depleted of Fe, Si, and Mg; O, and S are moderately depleted, while Ne displays a larger abundance. If we assume that soft X-rays are produced by plasma heated in the accretion process, the observed abundances probe the chemical composition of the infalling circumstellar material. We compared the present abundance values of MP~Mus with those obtained for the other three CTTSs showing evidence of X-ray emission due to shock-heated plasma (Table~\ref{tab:ctts}). In all cases the X-ray spectra indicate that the accreted material has a Ne abundance enhanced with respect to the other elements (with respect to the solar photospheric abundance ratios).

\citet{StelzerSchmitt2004} suggested that the non-solar abundances of the shock-heated plasma of TW~Hya may be explained by assuming that the accreting material underwent grain depletion, a mechanism already invoked by \citet{HerczegLinsky2002} to explain the low Si abundance. Depending on the temperature, the circumstellar material is composed by gas and dust phases which have different chemical compositions. The actual abundance ratios in the gas phase are determined by the different condensation temperatures $T_{\rm c}$ of the various elements \citep{SavageSembach1996}. One of the proposed scenarios of disk stratification predicts that dust grains mostly settle in the disk midplane, while the gas extends up to the disk surface, where it is easily ionized by the stellar radiation. Here it is funnelled along the magnetic field lines and accretes onto the central star. If the gas and dust phases are spatially separated in the inner circumstellar disk and/or the gas accretes more efficiently than the dust, the shock-heated material should be depleted of those elements (like Fe) which easily condense into dust grains and conversely enriched of more volatile elements (like noble gases). Following this scenario, the accreted material should have an abundance pattern similar to that observed in the interstellar gas, i.e. with the abundances decreasing for increasing $T_{\rm c}$. Note that the phenomenon of gas/dust separation in circumstellar disks, and the subsequent accretion of only the gas phase, is also held responsible for the peculiar abundances of RV~Tauri stars \citep[see][and references therein]{GiridharLambert2005}.

\begin{table}
\caption{CTTSs properties.}
\label{tab:ctts}
\begin{center}
\begin{tabular}{lcrlrr}
\hline\hline
Star      & \multicolumn{1}{c}{Stellar}         & \multicolumn{1}{c}{Age}   & \multicolumn{1}{c}{Ne/O} & \multicolumn{1}{c}{$T_{\rm med}^b$} & \multicolumn{1}{c}{$N_{\rm e}$} \\
          & \multicolumn{1}{c}{Association$^a$} & \multicolumn{1}{c}{(Myr)} &                          & \multicolumn{1}{c}{(MK)}              & \multicolumn{1}{c}{${\rm (10^{11}\,cm^{-3})}$} \\
\hline
BP~Tau    & Taurus & 0.6$^c$  & 0.43$^d$ & 16.1$^d$             & 3.2$^d$$\hspace{1.5em}$     \\
TW~Hya    & TWA    & 8$^e$    & 0.87$^f$ &  4.8$^d$             & 21.1$^d$$\hspace{1.5em}$    \\
V4046~Sgr & BPMG   & 12$^g$   & 1.04$^h$ &  6.4$^h$             & 3.2$^h$$\hspace{1.5em}$     \\
MP~Mus    & LCC    & 17$^i$   & 0.46     &  8.5$\hspace{0.4em}$ & 5.0$\hspace{1.9em}$         \\
\hline
\end{tabular}
\end{center}
$^a$~TWA = TW~Hydrae association, BPMG = $\beta$~Pictoris Moving Group, LCC = Lower Centaurus Curx; $^b$~$T_{\rm med}$ is derived as $(\Sigma_{i}EM_{i}\,T_{i})/(\Sigma_{i}EM_{i})$; $^c$~\citet{GullbringHartman1998}; $^d$~\citet{RobradeSchmitt2006}; $^e$~\citet{MakarovFabricius2001}; $^f$~\citet{DrakeTesta2005b}; $^g$~\citet{ZuckermanSong2001}; $^h$~\citet[][$T_{\rm med}$ has been derived from the line fluxes]{GuntherLiefke2006}; $^i$~\citet{MamajekMeyer2002}.
\end{table}


\citet{DrakeTesta2005b} showed that TW~Hya displays a Ne/O ratio larger by a factor $\sim2$ than the uniform value observed in a vast sample of coronal X-ray sources \citep{DrakeTesta2005a}. Moreover, Ne/Fe abundance ratios as large as that of TW~Hya were observed only in very few cases of putatively coronal sources. These results support the hypothesis that soft X-ray radiation from TW~Hya is not produced by coronal plasma. \citeauthor{DrakeTesta2005b} suggest that the Ne/O ratio can be used as a criterion to identify X-ray sources where the emitting material in the accretion stream suffered  grain depletion. However, we note that the condensation temperature of O is quite low (180\,K), therefore the separation between gas and dust must occur at low temperature to produce significant O depletion in the accretion streams.

A large Ne/O abundance ratio is observed also in the X-ray spectrum of V4046~Sgr \citep{GuntherLiefke2006}, where high density hints again at X-rays from shock-heated plasma. Instead, both MP~Mus and BP~Tau have Ne/O ratios typical of stellar coronae. \citet{DrakeTesta2005b} explained the Ne/O ratio of BP~Tau, lower than that of TW~Hya, on the basis of the different evolutionary stages of their circumstellar disks. Since TW~Hya is significantly older than BP~Tau, it is conceivable that the dust/gas separation process, and the subsequent depletion of high $T_{\rm c}$ elements, is not visible in the latter case because these processes occur on a time scale longer than the age of BP~Tau ($\sim 0.6$\,Myr).

In Table~\ref{tab:ctts} we report the ages of the four CTTSs introduced above. We adopt an age of 17\,Myr for MP~Mus \citep{MamajekMeyer2002}, since it is compatible with the age of the LCC association. For the subsequent discussion, the absolute age of each CTTS is unimportant, while only the age sequence matters, whose reliability depends only on the correctness of the membership of these CTTSs to the relevant stellar associations. 

\begin{figure}
\resizebox{0.95\hsize}{!}{\includegraphics{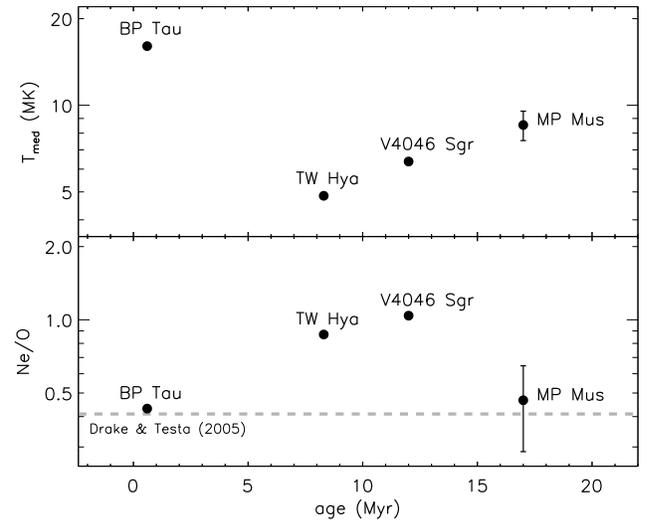}}
\caption{Average plasma temperature and Ne/O ratio vs age for the sample of four CTTSs with evidence of high density cool plasma.}
\label{fig:ctts}
\end{figure}

Figure~\ref{fig:ctts} shows the variations of plasma average temperature and Ne/O ratio with respect to stellar age, for the sample of four CTTSs (having spectral types ranging from K1 to K7). Both $T_{\rm med}$ and Ne/O do not have a monotonic trend with age, but these two plots suggest that stars with hotter plasma have lower Ne/O ratios, and vice versa. It is likely that high $T_{\rm med}$ indicates a large contribution from coronal plasma to the whole X-ray emission. In this scenario of mixed accretion-driven and coronal X-ray emission, the measured Ne/O ratio is a weighted average of the values in the shock-heated plasma and in the coronal plasma. Hence any large Ne/O ratio of the accreted material may be partly hidden by the coronal plasma abundances. To check this possibility we fitted the observed EPIC spectra of MP~Mus assuming a high Ne/O ratio for the coolest plasma component, but the model does not reproduce the observed spectra as well as the model described in Sect.~\ref{obs}. Moreover, in MP~Mus the hot coronal plasma does not contribute significantly to the observed O and Ne line emission (see below).

We conclude that the relatively low Ne/O ratio in MP~Mus is a characteristic of the cool accretion component, and the stellar age is likely not the only parameter which determines the Ne/O ratio observed in CTTSs with evidence of high density cool plasma.

\subsection{Accretion}

For the subsequent discussion we first assume that the cool X-ray emitting plasma of MP~Mus is only due to the shock accretion, with no contribution from coronal plasma. Starting from the \citet{MamajekMeyer2002} results on MP~Mus, and based on the \citet{SiessDufour2000} stellar models, we adopt for MP~Mus a mass of $1.2\,{\rm M_{\sun}}$ and a radius of $1.3\,{\rm R_{\sun}}$.

Using the \ion{O}{vii} triplet and the \ion{O}{viii} Ly$\alpha$ lines we infer the electron density ($N_{\rm e}=5\times10^{11}\,{\rm cm^{-3}}$), temperature ($T=3$\,MK, obtained from the \ion{O}{viii} Ly$\alpha$ and \ion{O}{vii} $r$ lines), and emission measure ($EM=2.4\times10^{53}\,{\rm cm^{-3}}$) of the post shock plasma. In the strong shock scenario, the relevant plasma parameters are linked by the relations:

\begin{equation}
N_{1}=4N_{0},\;\; v_{1}=\frac{1}{4}v_{0},\;\;
T_{1}=\frac{3}{16}\frac{\mu m_{\rm H}}{k}v_{0}^2 
\end{equation}

\noindent
where the suffixes $0$ and $1$ indicate the pre-shock and post-shock plasma, $N$ the density, $v$ the velocity, $T$ the temperature, and $\mu$ the mean molecular weight (in our case $\mu = 0.61$). From the measured temperature $T_{1}$ we infer that the pre-shock velocity is $470\,{\rm km\,s^{-1}}$. This value corresponds to a free fall from an inner radius of the circumstellar disk of $3\,R_{\star}$, or from a larger distance if some energy loss occurs during the fall. From the post-shock plasma temperature and density we derive a cooling time of 350\,s, and considering that the post-shock velocity is $120\,{\rm km\,s^{-1}}$, we obtain a characteristic length of the post-shock region $l = 4\times10^9\,{\rm cm}=0.05\,R_{\star}$. Hence, the cross section of the infalling stream $A=EM/(N_{\rm e}N_{\rm H}l)$ is $3\times10^{20}\,{\rm cm^{2}}$. It corresponds to a filling factor $f=A/(4\pi R_{\star}^{2})$ of $0.3\,\%$ of the stellar surface, and to a mass accretion rate of $5\times10^{-11}\,{\rm M_{\sun}\,yr^{-1}}$.

We made the hypotheses that: (1) the cool plasma is produced in the accretion shock; (2) the cool plasma is optically thin; (3) its density is measured from the \ion{O}{vii} $f/i$.

We are confident that the assumption (1) is appropriate. First note that the two strong flares detected, produced by coronal plasma, contribute just 3.6\% of the spectrum above 18\,\AA~(i.e. below 0.7\,keV); second, we tried to fit the \ion{O}{vii} triplet with two contributions due to low and high electron density ($10^{9}$ and $10^{12}\,{\rm cm^{-3}}$, respectively), finding that at least 80\% of the \ion{O}{vii} is due to high density (i.e. shock-heated) plasma, and at most 20\% to low density (i.e. coronal) plasma.

Hence the derived accretion rate {\it \.{M}}, which depends only on the hypothesis (1) and (2), but not on $N_{\rm e}$, is acceptable, but a larger {\it \.{M}} could be possible if part of the X-ray emission is absorbed.

The measured $N_{\rm e}$ is more uncertain: a small contribution of low $N_{\rm e}$ coronal plasma to the \ion{O}{vii} triplet might cause an underestimation of $N_{\rm e}$; conversely an UV field might influence the populations of the \ion{O}{vii} atomic levels by photoexcitation and hence mimic an high density plasma. 
No UV excess emission, which could originate from the accretion hot spot, has been reported for this star. However a sufficiently high UV radiation density can be present only very near the accretion hot spot on the stellar surface, and the photoexcitation hypothesis would anyway indicate that the cool X-ray emitting plasma is close to the base of the accretion funnel.

\section{Conclusions}

From the analysis of the {\it XMM-Newton} observation of the CTTS MP~Mus we derived evidences that plasma heated in the accretion shock produces the soft part of the X-ray emission. We measured a Ne/O ratio similar to that of BP~Tau and reduced by a factor 2 with respect to that of TW~Hya and V4046~Sgr: this result suggests that the stellar age is not a useful parameter to predict the amount of grain depletion suffered by the accreting material.

\begin{acknowledgements}

CA, AM, and GP acknowledge partial support for this work from contract ASI-INAF I/023/05/0 and from the Ministero dell' Universit\`a e della Ricerca. Based on observations obtained with {\it XMM-Newton}, an ESA science mission with instruments and contributions directly funded by ESA Member States and NASA.

\end{acknowledgements}

\bibliographystyle{aa} 
\bibliography{mpmus}

\end{document}